# Analysis of the use of smart cards on the urban railway

Dmitry Namiot[1], Oleg Pokusaev[2], Vasily Kupriyanovsky[3]

[1]Lomonosov Moscow State University, [2]Moscow; Center of digital high-speed transport systems Russian University of Transport (MIIT), Moscow; [3]National Competence Center for Digital Economy Lomonosov Moscow State University, Moscow

dnamiot@gmail.com  o.pokusaev@rut.digital  vpkupriyanovsky@gmail.com

**Abstract.** The article analyzes the patterns of use of railway stations in the Moscow region. The basis for the analysis is the data of smart cards on the entrances and exits of passengers. The technical tool is time series similarity analysis. As a result, the work identifies the main patterns of user behavior on the use of railway stations (railway transport). The results of the work were used in the design of new lines of urban railways. Obviously, the use patterns reflect the current state of the transport system and the urban environment. Accordingly, the recorded changes in usage patterns can serve as indicators and metrics for changes in the urban environment.

## 1 Introduction

In our work, we investigated the patterns of the use of railway stations in the city and the suburbs. The first question for any new transport project is always the prediction of traffic (the estimation of use) of the new transport system. Actually, the traffic defines all economic and social aspects of any transport project. In order to provide such predictions, we need to understand the patterns of the movement of passengers (the models of using the transport system). The existing patterns and their possible changes are the base for our prediction. Identification of such patterns and their use in urban analytics are the main subjects of this paper. Obviously, the transport use patterns reflect the current state of the transport system and the urban environment. Accordingly, the discovered changes in usage patterns can serve as indicators and metrics for changes in the urban environment. It is how the transport behavior patterns could be used in urban analytics.

Mobility (or smart mobility) is one of the key characteristics of the Smart City according to all standards. Accordingly, all cities are constantly involved in the development of transport projects. Cities are no longer designed for cars. Modern cities are oriented, rather, to pedestrians (so-called pedestrian economy) [1]. A convenient opportunity to use several modes of transport is the basic requirement for smart transport in a smart city. Mobility services should be multimodal. Intellectual mobility is defined as the use of technology and data to create links between people, places, and goods in all modes of transport. Mobility as a service is a new concept that offers consumers access to various types of vehicles and experience of travel [2]. Mobility as a service can be perceived as a "better choice" for organizing a move, and this can change the way we are currently treating transport. The key issue for creating applications of the class "mobility as a service" is the availability of digital information from various sources of the urban economy [3].

Recently, all cities in the world show interest in rail transport development (urban railways). This is due to several and quite natural causes that are common to all cities. Railway transport (the urban railway) for today, for example, is the only way to organize traffic without traffic jams.

Note also that the very concept of the city has expanded significantly. Today, in most cases, we should talk about agglomerations, since a large number of people living outside the city constantly go to the city to work, study, etc.

It is obvious that unfortunately, all transport projects are quite expensive. And the first part of any such project always includes an estimate of possible traffic. As we pointed out above, this is based on the understanding of the models of the use of the transport system in the city. For example, passengers, in reality, cannot always choose the shortest route, if it is connected, for example, with extra movement. The choice will be affected by the possible time savings and comfort of a change [4].

In this paper, we present the results of an analysis of the use of railway stations, which were, in fact, aimed at identifying patterns (models) for using the urban railroad. This was a part of the project to build new lines of urban railways in Moscow (Fig. 1).

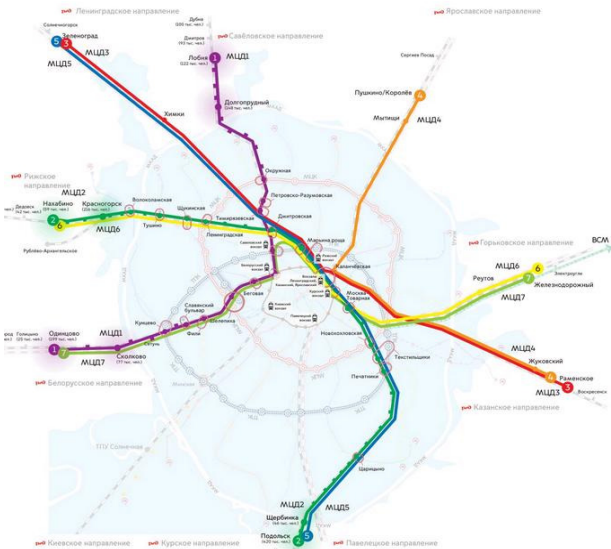

**Figure 1**. Moscow railways diameters

The rest of the paper is organized as follows. In Section 2, we describe the available data. Section 3 presents our work on the analysis of transport behavior models. In Section 4, we discuss usage patterns. In Section 5, we discuss related works.

## 2 On available data

The current model of using passenger rail transport in Moscow and the region assumes that every passenger validates his travel document (ticket) at least twice - when boarding and leaving a station. In terms of social networks, these are two marks - check-in and check-out [5]. This makes the information on the use of railway stations very useful and, in a sense, a unique compared with other sources of data on movements. From this information, we can immediately restore the route (by rail, of course). Traditionally, travel documents (smart cards, for example) in Moscow are validated only at the entrance (check-in). Accordingly, to restore the route, we must use heuristic algorithms [6].

For example, if we noted the use of a travel card (in Moscow, for example, it is so-called Troyka card) at point A, and then after a relatively long break at point B, then we can assume that the passenger traveled from A to B, then, for example, was at work and after its completion makes (starts) a new trip at B. The allocation of routes based on the data of telecommunication operators is also relied on heuristics [7]. For example, the place where calls are made in the morning and in the late evening is considered "home", in the daytime - by "work" and so on. In the case of information on the use of railway stations, the route is known [8].

The data for railways stations entrances and exits are presented as separate files (CSV), each of which describes the passes for a particular station in one month. One entry (a line in the file) corresponds to one pass (to the entrance or to the exit). The data is completely anonymous. The type of document used (for example, preferential or not, one-time or re-used) is present for each record, but there is no identification of documents at all.

The size of each such file depends, of course, on the use of a particular station in a particular month and varies between 20 and 70 Mb.

Fields that are contained in the records:

- Date and time
- Characteristics of the price (full or discounted ticket)
- Type of benefits for discounted tickets (Federal, Russian Railways, etc.)
- Type of ticket (one-way ticket, round trip, one-time and one-way ticket, subscription, etc.)
- The information carrier (paper ticket or smart card)
- Starting Station
- End station

To analyze the data, we've used a cloud tool from Google - Collaboratory. Colaboratory is a Google free to use research project created to help disseminate machine learning education and research. Technically, it is a Jupyter notebook environment that requires no setup for use and runs entirely in the cloud. Data for processing are stored in Google Drive [9].

## 3. On our model

The main idea of our research is to analyze the use of the railway station in time. In other words, it is not just some total number of passengers that enters and leaves the station during the day. The interesting thing is how incoming (departing) passengers are allocated in time. This information determines the mode of operation of the station. Also, this information helps to estimate the potential changes in the flow of passengers. The last but also important reason is the conclusion that this information helps to distinguish (classify) railways stations. As it will be shown below, the actual use of Moscow agglomeration stations differs from the standard (generally accepted pattern) - peak in the morning, when passengers go to work and a second peak in the evening when they return back home. The real picture is much more diverse and it is different for various groups of stations.

To describe the distribution of the entries (exits) of a particular station, we used time-aggregated information about the validations of travel documents. For example, we can aggregate data for 60 or 30 minutes time interval. For further consideration, it is important that we can always use the same aggregation period for all data. Such aggregated data is a typical example of time series - the time of day and the number of passengers classified by this time. We can talk about the template if we find that such time series for a particular station are similar to each other. That is, we can compare such series for Mondays within a month, for Tuesdays, etc.

Actually, the first result that was obtained (and which determines the possibility of all further reasoning) is that railway passengers demonstrate enviable constancy. For each selected station, all working days are similar to each other (according to traffic characteristics). The same applies to weekends. The description of the distributions is exactly the pattern for the use of the railway station. For some stations, these templates are almost the same, for some stations they are different. This difference is explained by the location of the station, the density and composition of the population in the vicinity of the station, the presence of centers of attraction.

Technically, the search for templates is reduced to determining the similarity of time series. So, the main tool here is a time series similarity measurement [10]. Given two time series $T_1$ and $T_2$, a similarity function calculates the distance between the two time series. In our case, we will refer to distance measures that compare the $i$–th point of one time series ($T_1$) to the $i$–th point of another ($T_2$).

## 4 On railway stations use in Moscow agglomeration

The idea of data analysis consists of two main points. The patterns of movement that will be extracted from the data (can be found in the data) are a reflection of some existing socio-economic processes in the urban agglomeration. Where do the workers live, where are their jobs in the city, what mode of work, etc. Accordingly, the conclusions that will be made on the basis of data analysis must have some explanation from the point of view of these processes. Let's call this "urban" explanation. And vice versa. Some changes in the observed data can serve as an indicator (or even a metric) of changes in the city (in the agglomeration area).

As far as we know, earlier the analysis of information about the arrivals and departures was not carried out. We can say that the data was processed with only an accounting goal: the number of people who passed and the total accumulated revenue.

The first thing we wanted to investigate is the mode of using the road. According to the general ideas, people go to work in the morning and return back in the evening. Accordingly, we should see a peak at the entrances in the morning at some stations, then (with a delay for the duration of the trip) - the peak in the number of exits at other stations. In the evening, the picture should be reversed.

Here are the hourly figures of the passes for the station outside Moscow. The upper part (Fig. 3) presents entrances (departures from the station) and the lower one (Fig. 4) presents exits (arrivals).

For all images below, the X-axis represents the time, the Y-axis represents the number of passengers entering or leaving a particular station. All the graphs for a particular station represent an average pattern for using this station on a working day or a weekend. Before the presentation of this template, the time series similarity for different days of the week was checked (Section 3).

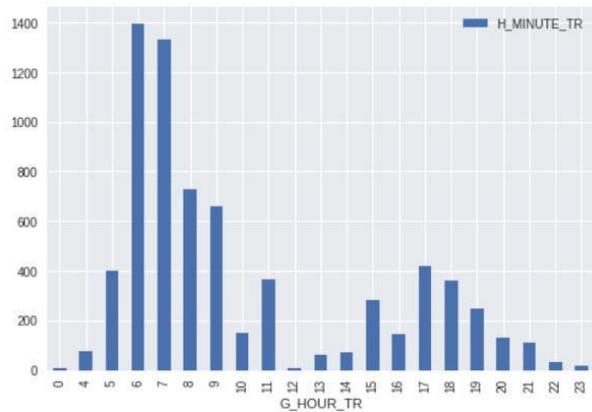

**Figure 2**. Outside of Moscow entrances

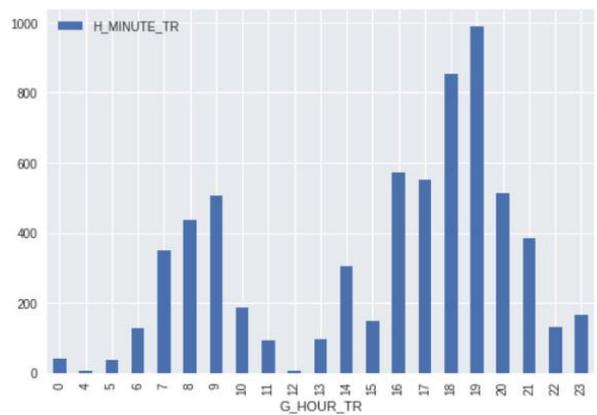

**Figure 3.** Outside of Moscow exits

The picture seems to meet expectations. At 6-7 am, passengers are sent to work (peaks on the first graph - entrances to the station), at 6-7pm hours - they are returned (peaks on the second graph are exits). But let us see the same graphs for the station inside Moscow.

At the entrances, we already have two peaks - in the morning and in the evening. In the morning, people go to work (the station is used as a subway station), in the evening - they leave this station on the way home to the suburb. The morning peak is shifted relative to the "regional" for one hour – people are closer to work and can leave home later. The evening peak is at the same time, since the end time for work is the same.

Note the very low level at the middle of the day (1-2 pm). Partly, it is linked to so-called technological break (lack of trains). Immediately after the break, there is quite a lot of demand. For stations outside the city, we do not see such demand. This suggests, that the proposed cancellation of a break on new lines will be highly demanded within the city. Technically, without this interruption, the city rail line will function as a metro line with the regular intervals between trains.

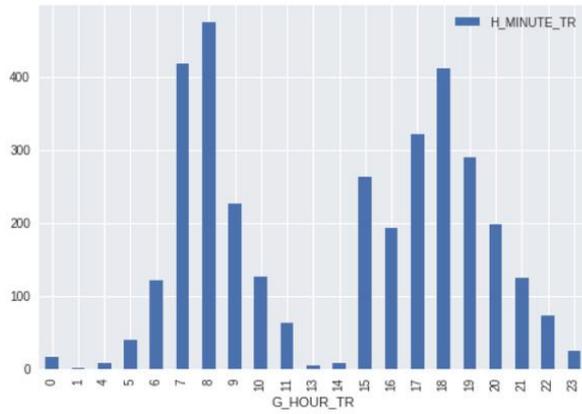

**Figure 4**. Entrances inside of Moscow

On working days of the week, these distributions remain stable. But on weekends the picture changes. People from the regional station continue to leave for work in the morning, but also they go to the city during the day, apparently already with some private purposes (Fig. 5).

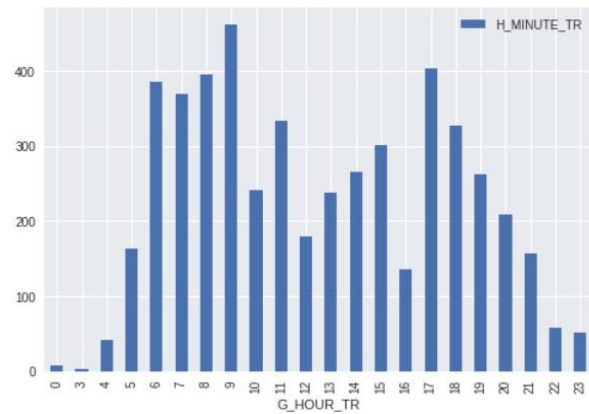

**Figure 5.** Weekends entrances outside of Moscow

The departures last till 7-9pm hours. And here is the picture of the exits (arrivals to the station):

Morning peaks on this chart are, apparently, summer residents or those who live on weekdays in the city and go home for the weekend.

At the station inside of Moscow, we see the morning traffic, which is determined by the people working at the weekend. It is interesting that the number of passengers is practically the same as on working days. Also, there are no daily dips - people continue to drive, unlike working days.

For stations inside of Moscow, there is no evening peak at the weekend. In other words, those working on weekends these days do not return home through this station, since they did this on weekdays. One possible explanation is the use of cars. Alternatively, these people (their offices, businesses, for example, in the vicinity of the station) do not work on weekends. So, at the weekends, the stations inside of Moscow work more for passengers from Moscow. The flow of passengers to the suburb area (especially in the evening) decreases noticeably.

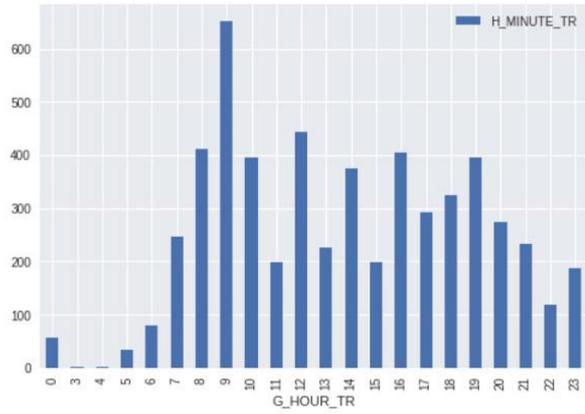

**Figure 6.** Weekends exits outside of Moscow

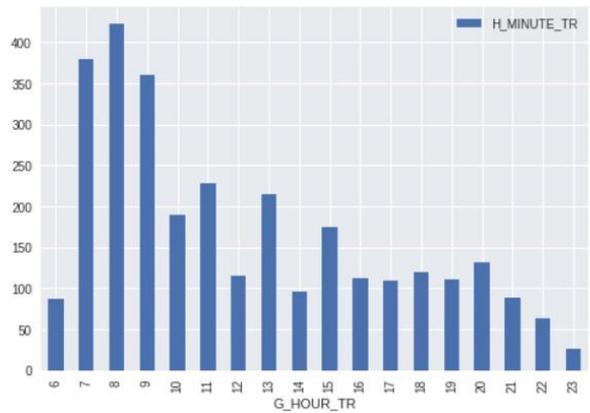

**Figure 7.** Weekends entrances (departures) inside of Moscow

## 5 Related works

The review of algorithms for analysis and recovery of routes based on validation of travel cards is contained, for example, in our paper [6].

In the paper [11], the authors use smart card records resolved in both time and space for getting collective spatial and temporal mobility patterns at large scales and reveal the regularity of these patterns. The main goals declared there are very close to our contribution: demonstrate the potential of using smart card records as data sources to gain insights into city dynamics and aggregated human behavior; explore the relationship between spatiotemporal patterns of smart card usage and underlying city behavior and geography; study patterns in smart card usage, including an analysis of how factors such as the time of the day affect this prediction. However, it should be noted that the work is trying to identify the individual patterns of displacement too. In our case, this is impossible, since there is no information on the identification of travel documents.

In the paper [12], the authors target the fundamental dilemma: how to make an urban development less dependent upon mobility by car. Examples of locating transit patterns, including extensive literature analysis, are contained in the paper [13]. In the paper [14], authors describe check-in/check-out processing models for traffic prediction. Smart cards (travel cards) present the main source for data mining and predictions. The typical example is paper [15]. Authors provide in-depth temporal and spatial analysis for individual travel patterns, analyze the relationship between temporal and spatial features, and perform abnormal detection. Another good example is the paper [16]. The main idea here is to detect clusters for passengers with the similar transport behavior.

## 6 Conclusion

The paper discusses the use of data on the use of railway stations in the Moscow agglomeration as a tool for analyzing transport behavior. The results outlined in this article have found practical application in the works on designing a new system of urban railways in the Moscow region. The paper investigated the relationship of the results of processing data on the use of railway stations with socio-economic aspects of the life of the inhabitants of the Moscow agglomeration. The authors propose the method for getting usage patterns for railway stations. This construction (allocation) of usage patterns of stations connected with work traffic is considered.